# Relativistic Spectrum of Hydrogen Atom in Space-Time Non-Commutativity

Mustafa Moumni[a], Achor BenSlama[b] and Slimane Zaim[c]

[a]*Matter Sciences Department, Faculty of SE&SNV, University of Biskra, Algeria*
m.moumni@univ-biskra.dz
[b]*Physics Department, Faculty of Sciences, University of Constantine, Algeria*
a.benslama@yahoo.fr
[c]*Matter Sciences Department, Faculty of Sciences, University of Batna, Algeria*
slimane69zaim@yahoo.com

**Abstract.** We study space-time non-commutativity applied to the hydrogen atom via the Seiberg-Witten map and its phenomenological effects. We find that it modifies the Coulomb potential in the Hamiltonian and add an $r^{-3}$ part. By calculating the energies from Dirac equation using perturbation theory, we study the modifications to the hydrogen spectrum. We find that it removes the degeneracy with respect to the total angular momentum quantum number and acts like a Lamb shift. Comparing the results with experimental values from spectroscopy, we get a new bound for the space-time non-commutative parameter.

## INTRODUCTION

The idea of taking non-commutative space-time coordinates is not new as it dates from the thirties of last century. It had as objective to regulate the divergences of quantum field theory by introducing an effective cut-off coming from a non-commutative structure of space-time at small length scales. Then, it was abandoned because of problems caused by the violation of unitarity and causality, but the mathematical development of the theory continued and especially after the work of Connes in the eighties of the same century [1].

In 1999 and during their work on string theory, Seiberg and Witten showed that the dynamics of the endpoints of an open string on a *D*-brane in the presence of a magnetic back-ground field is described by a theory of Yang-Mills on a non-commutative space-time [2]; this has renewed interest in the theory. Recently, there are many works on Lorentz invariant interpretation of the theory [3-5].

The theory is a distortion of space-time where the coordinates $x^\mu$ become Hermitian operators and thus do not commute:

$$\left[ x_{nc}^\mu, x_{nc}^\nu \right] = i\theta^{\mu\nu} = i\,C^{\mu\nu}/\Lambda_{nc}^2 \,;\, \mu,\nu = 0,1,2,3 \qquad (1)$$

$\theta^{\mu\nu}$ is the deformation parameter and *nc* indices denote non-commutative coordinates. The $C^{\mu\nu}$ are dimensionless parameters and $\Lambda$ is the energy scale where the non-commutative effects will be relevant. The non-commutative parameter $\theta^{\mu\nu}$ is an anti-symmetric real matrix and ordinary space-time is obtained by making the limit $\theta \to 0$. For a review, one can see reference [6] and a well-documented review on bounds of the non-commutative parameter can be found in reference [7].

In this paper, we are interested in the phenomenological consequences of space-time non-commutativity. We focus on the hydrogen atom because it is a simple and a well-studied quantum system and so it can be taken as an excellent test for non-commutative signatures. The space-space case was studied in the non-relativistic case in [8] and [9], while relativistic treatment was done in [10]. For the space-time case, only the non-relativistic theory was studied in both [10] and [11]. Here we work on the space-time case in relativistic theory. We start by computing the corrections to the Dirac energies, then we study the effect on transitions energies and finally we comparing with experimental results from hydrogen spectroscopy.

## HYDROGEN ATOM IN SPACE-TIME NON-COMMUTATIVITY

To build the non-commutative image of a commutative theory, it suffices to use the Seiberg-Witten map whose main ingredient is the star product [2]:

$$(f*g)(x) = exp\left(\frac{i}{2}\theta^{\mu\nu}\frac{\partial}{\partial x^\mu}\frac{\partial}{\partial y^\nu}\right)f(x)g(y)\Big|_{y=x} = f \cdot g + \frac{i}{2}\theta^{\mu\nu}\partial_\mu f \partial_\nu g + O(\theta^2) \qquad (2)$$

Here, we restrict ourselves to the 1st order in the deformation parameter because of its smallness [7]. For the $U(1)$ gauge theory one take the action defined by:

$$S_{(nc)} = -\tfrac{1}{4}\int d^4 x F_{\mu\nu} * F^{\mu\nu} = -\tfrac{1}{4}\int d^4 x \hat{F}_{\mu\nu}\hat{F}^{\mu\nu} \qquad (3)$$

where the noncommutative fields' strengths come from the relations:

$$\hat{F}_{\mu\nu} = \partial_\mu \hat{A}_\nu - \partial_\nu \hat{A}_\mu - ie\left[\hat{A}_\mu, \hat{A}_\nu\right]_* \qquad (4)$$

and write noncommutative fields as a series expansion according to powers of $\theta$.

By solving the equations of the Seiberg-Witten map to the 1st order in the deformation parameter, one can found:

$$\begin{cases} \hat{A}_\mu = A_\mu + A_\mu^{(1)} + O(\theta^2) \\ \hat{F}_{\mu\nu} = F_{\mu\nu} + F_{\mu\nu}^{(1)} + O(\theta^2) \end{cases} \Rightarrow \begin{cases} A_\mu^{(1)} = -(1/2)\theta^{\alpha\beta} A_\alpha\left(\partial_\beta A_\mu + F_{\beta\mu}\right) \\ F_{\mu\nu}^{(1)} = \theta^{\alpha\beta}\left(F_{\mu\alpha}F_{\nu\beta} - A_\alpha \partial_\beta F_{\mu\nu}\right) \end{cases} \qquad (5)$$

Where *A* and *F* are the usual fields of the ordinary $U(1)$ theory (for details, one can see [9] for example).

Applying this procedure to the Coulomb potential when non-commutativity is between time and space coordinates, we get (in natural units $\hbar=c=4\pi\varepsilon_0=1$):

$$eA_0^{(nc)} = -\frac{e^2}{r}\left(1+\frac{e^2}{r^3}\theta^{0j}x_j\right)+O(\theta^2) = -\frac{e^2}{r}-\frac{e^4}{r^3}\frac{\vec{\theta}\cdot\vec{r}}{r}+O(\theta^2) \quad (6)$$

The *j* indices correspond to spatial coordinates while 0 indicates time and we have used the vectorial notation $\vec{\theta} = (\theta^{01},\theta^{02},\theta^{03}) = (\theta^1,\theta^2,\theta^3)$.

## Non-Commutative Corrections of Dirac Energies

To compute the corrections induced by non-commutativity to relativistic energies, we write the ordinary Dirac equation but with the deformed Coulomb potential (6):

$$i\hbar\partial_0\psi = H\psi = \left(\vec{\alpha}\cdot\vec{p}+m\gamma_0+eA_0^{(nc)}\right)\psi \quad (7)$$

and we handle the additional term proportional to *θ*, with perturbation method; so the corrections to eigenvalues or ordinary Dirac energies are given by the mean values:

$$\langle\psi|\Delta H^{(nc)}|\psi\rangle = \langle e\Delta A_0^{(nc)}\rangle = \left\langle-\frac{e^4}{r^3}\frac{\vec{\theta}\cdot\vec{r}}{r}\right\rangle \quad (8)$$

We work in spherical coordinates and make the choice $\vec{\theta} = \theta^{0r}\vec{r}/r$ (it is identical to that made in [12] and [13]), so we get:

$$\langle\Delta H^{(nc)}\rangle = \left\langle-\frac{e^4}{r^3}\frac{\theta^{0r}r}{r}\right\rangle = -e^4\theta\langle r^{-3}\rangle \quad (9)$$

The mean value is computed using the Eigen functions of Dirac equation for the usual Coulomb potential. It is not difficult to do this using the solutions of the relativistic equation (see [14] for example), or we can directly use recurrence relations from [15]:

$$\left\langle\frac{1}{r^3}\right\rangle = 2(1-\epsilon^2)^{3/2}\frac{3\epsilon^2\kappa^2-3\epsilon\kappa-\kappa^2+\alpha^2+1}{\sqrt{\kappa^2-\alpha^2}(\kappa^2-\alpha^2-1)(4\kappa^2-4\alpha^2-1)}(m)^3 \quad (10)$$

In this expression, *α* is the fine structure constant, $\epsilon = E/m$ and $\kappa=j+1/2$ if $j=l-1/2$ or $\kappa=-(j+1/2)$ if $j=l+1/2$. *E* represents the habitual Dirac energy for the Hydrogen atom.

We see that through *κ*, the energy depends not only on the value of *j* but also on the manner to get this value (or on *l*), unlike the usual Dirac energies which are the same for all the possible ways to obtain *j*. It implies that the non-commutativity removes the degeneracy $j=l+1/2=(l+1)-1/2$ in H-atom (states like $nP_{3/2}$ and $nD_{3/2}$ for example); we say here that non-commutativity has an effect similar to the Lamb shift.

**TABLE 1.** Non-Commutative corrections to the levels $n=1,2,3$ of H-atom ($\theta$ is in $eV^{-2}$)

| Level | N.C Correction ($eV$) | Level | N.C Correction ($eV$) | Level | N.C Correction ($eV$) |
|---|---|---|---|---|---|
| $1S_{1/2}$ | $-2.074\times10^{11}\,\theta$ | $2P_{1/2}$ | $-1.438\times10^{5}\,\theta$ | $3P_{3/2}$ | $3.409\times10^{4}\,\theta$ |
| $2S_{1/2}$ | $-2.593\times10^{10}\,\theta$ | $2P_{3/2}$ | $1.151\times10^{5}\,\theta$ | $3D_{3/2}$ | $6.8179\times10^{3}\,\theta$ |
| $3S_{1/2}$ | $-7.683\times10^{9}\,\theta$ | $3P_{1/2}$ | $-5.682\times10^{4}\,\theta$ | $3D_{5/2}$ | $6.8177\times10^{3}\,\theta$ |

To put the corrections to the Lamb shift in a more visible and general form, we write the expressions by using the development in series with respect to $\alpha^2$ and we will stop at the first non-vanishing order, because the next term will be $10^{-5}$ smaller ($\alpha^2=5\cdot10^{-5}$):

$$\Delta E_{n,j}^{(nc)}(LambShift) = E_{n,j=l+1/2}^{(nc)} - E_{n,j=(l+1)-1/2}^{(nc)} = \frac{12 m^3 e^4 \theta}{j(j+1)(2j-1)(2j+3)} \frac{\alpha^3}{n^3} + O(\alpha^5) \quad (11)$$

Note that this approximated relation is divergent only for $j=1/2$ and we have to use the exact expressions from (9) and (10) for this case because it gives us a finite result as one can see in TABLE 1 (the Kramer's relation $<r^{-n}>$ in the non-relativistic case are also divergent for $l=0$ for all values $n>2$).

We can get a limit for $\theta$ by comparing these shifts to experimental results from Hydrogen spectroscopy. We take as an example the transition $2P_{1/2}\rightarrow 2S_{1/2}$ (the 28cm line or Lyman-$\alpha$). From TABLE 1, we have ($\theta$ is in $eV^{-2}$):

$$\Delta E_{2,1/2}^{(nc)}(LambShift) = \Delta E_{2P_{1/2}\rightarrow 2S_{1/2}}^{(nc)} = 2.593\times10^{10}\,\theta\,eV^3 \quad (12)$$

We compare this result to the current experimental accuracy 2.4 $kHz$ from [16] and find the bound $\theta_{st}<(90\ GeV)^{-2}$. This new limit is better than those obtained in both [9] and [11]. We can use the $2S\rightarrow 1S$ transition and get a better value because the frequency of this transition has the most accurate experimental precision (S states have a large natural lifetime and because of that, their spectral lines are narrow and so the precision is better). The accuracy for this transition is 34 $Hz$ from [17] and the limit obtained is with this value is $\theta_{st}<(2.0\ TeV)^{-2}$.

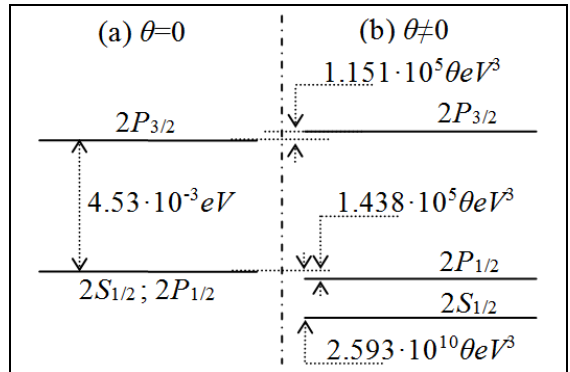

**FIGURE 1.** Relativistic corrections from space-time case of non-commutativity to $n=2$ level of H-atom in (b) are compared to the usual relativistic shift coming from habitual Dirac theory in (a). (The unit of the parameter $\theta$ is $eV^{-2}$)

# CONCLUSION

In this work, we have studied space-time non-commutative hydrogen atom and induced phenomenological effects. We found that applying space-time non-commutativity to the electron in the hydrogen atom modifies the Coulomb potential so we get an additional term proportional to $r^{-3}$. By computing the corrections induced by this additional term to the Dirac theory of H-atom, we found that space-time non-commutativity removes the degeneracy of the Dirac energies with respect to the total angular momentum quantum number $j=l+1/2=(l+1)-1/2$, and the new energies are labeled $E_{njl}$.

By similarity, we have concluded that the non-commutativity has an effect similar to that of the Lamb Shift. This is explained by the fact that Lamb correction can be interpreted as a shift of *r* due to interactions of the bound electron with the fluctuating vacuum electric field [18] and non-commutativity is also a shift of *r* as one can see from (6) (another formulation of non-commutativity in a shift of *r* called the Bopp's shift [19]).

By comparing the corrections found here to experimental results from high precision hydrogen spectroscopy, we get a new bound for the parameter of non-commutativity around $(2.0\ TeV)^{-2}$; this limit is an improvement of precedent bounds.

# ACKNOWLEDGMENTS

The authors would like to thank all the members of the LPMPS from the University of Constantine for their help and availability during the realization of this work.

This work was carried out within the National Research Programme in Algeria PNR (8/P/461).

# REFERENCES

1. A. Connes, *Noncommutative Geometry*, San Diego CA: Academic Press, 1994.
2. N. Seiberg and E. Witten, *JHEP* **9909**, 032, (1999).
3. M. Chaichian, P. Kulish, K. Nishijima and A. Tureanu, *Phys. Letters.* **B604**, 98–102 (2004).
4. C. D. Carone and H. J. Kwee, *Phys. Rev*. **D73**, 096005 (2006).
5. S. Saxell, *Phys. Letters*. **B666**, 486–490 (2008).
6. R. J. Szabo, *Phys. Rept.* **378**, 207-299 (2003).
7. R .J. Szabo, *Gen. Relativ. Gravit*. **42**, 1-29 (2010).
8. M. Chaichian, M. M. Sheikh-Jabbari and A. Tureanu, *Phys. Rev. Letters*. **86**, 2716 (2001)
9. A. Stern, *Phys. Rev.* **D78**, 065006 (2008).
10 T. C. Adorno, M. C. Baldiotti, M. Chaichian, D. M. Gitman and A. Tureanu, *Phys. Letters*. **B682**, 235-239 (2009).
11 M. Moumni, A. BenSlama and S. Zaim, *J. Geom. Phys.* **61**, 151-156 (2011).
12 S. Fabi, B. Harms and A. Stern, *Phys. Rev*. **D78**, 065037 (2008).
13 M. Chaichian, A. Tureanu, M. R. Setare and G. Zet, *JHEP* **0804**, 064 (2008).
14 H. Bethe and E. Salpeter, *Quantum Mechanics of One- and Two-Electron Atoms*, Berlin: Academic Press, 1957.
15 S. K. Suslov and, B. Trey, *J. Math. Phys*. **49**, 012104 (2008).
16 National Institute of Standards and Technology http://physics.nist.gov/PhysRefData/HDEL/transfreq
17 T. W. Hänsch et al, *Phil. Trans. Royal. Soc.* **A363**, 2155-2163 (2005).
18 C. Itzykson and J. B. Zuber, *Quantum Field Theory*, Mineola, NY: Dover Publications Inc (1980).
19 S. Dulat and K. Li, *Eur. Phys. J.* **C54**, 333-337 (2008).